# MICE: THE INTERNATIONAL MUON IONISATION COOLING EXPERIMENT

P. Drumm[†], CCLRC Rutherford Appleton Laboratory, Didcot, UK.

*Abstract*

Muon storage rings have been proposed for use as sources of intense high-energy neutrino beams and as the basis for muon colliders. Phase-space compression (cooling) of the muon beam prior to acceleration and storage is needed to optimise performance and cost. Traditional techniques cannot be employed to cool the beam because of decay during the short muon lifetime. Ionisation cooling, a process in which the muon beam is passed through an alternating series of liquid-hydrogen absorbers and accelerating RF-cavities, is the technique proposed to cool the muon beam. An international collaboration has been formed to carry out the Muon Ionisation Cooling Experiment (MICE), and its proposal to Rutherford Appleton Laboratory (RAL) has been approved. The status of the MICE cooling channel, the instrumentation, and the implementation at RAL are described, together with the predicted performance of the channel and the measurements that will be made.

## INTRODUCTION

Ionisation cooling should work, at least in theory. As an example, the neutrino factory (NF) design study-II [1] included more than 100 m of cooling channel of alternating liquid-hydrogen absorbers and RF acceleration. The estimated performance of this particular channel was to reduce the transverse beam emittance from 12 to ~2.7 mm-rad. The cost of the channel was estimated at ~20% of the total accelerator costs including the driver [1].

### The design aim

The aim of MICE is to show that one can engineer and build a section of a real cooling channel and place it into a real beam to understand the measured behaviour and extrapolate to a full NF cooling channel. It is difficult to reproduce exactly the beam of a NF from a conventional source because of various correlations that would be present in a real NF beam, but the important features of a cooling channel can be reproduced: large emittance can be generated with a high Z diffuser, and a matched beam of desired energy spread can be arranged with the muon beam optics.

### Performance and measurement technique

The effect on a muon beam of a short section of cooling channel (such as that of study-II) is shown in figure 1 for the transverse emittance (full transmission for a normalised emittance of less than ~6π mm-rad). The cooling ($\varepsilon_{out}/\varepsilon_{in}-1$) is predicted to be ≈10% and requires a measurement of sufficient accuracy to eliminate systematic and statistical uncertainties; an accuracy of $10^{-3}$ should be possible in $\varepsilon_{out}/\varepsilon_{in}$. To achieve this, conventional beam emittance measuring techniques have been replaced by direct measurement of individual particles as they enter and exit the MICE channel, from which the collective emittance of a beam can be reconstructed [2].

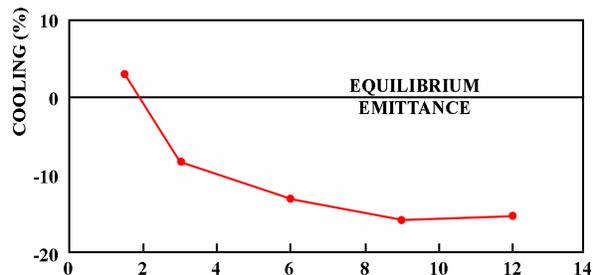

Figure 1. Cooling effect as a function of $\varepsilon_{n.}$, the normalised input beam emittance (π mm-rad).

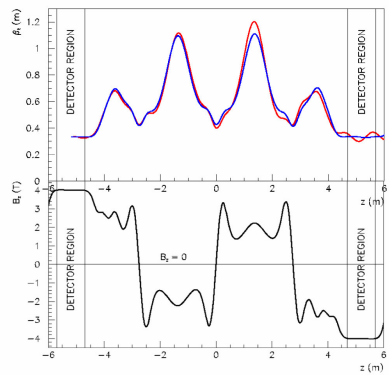

Figure 2. Anti-symmetric magnetic field profile in the MICE channel.

## THE COOLING CHANNEL

### The cooling channel design

The MICE cooling channel and instrumentation is shown in the 3D-cutaway drawing of figure 3. The cooling channel of MICE consists of three absorbers and two RF-cavity modules. The absorbers are surrounded by a pair of focusing magnets to provide the small beta-function in the absorbers and the RF cavities are surrounded by large diameter "coupling coils" that guide the beam within the cavities. Figure 2 also shows the baseline channel optics. This channel has the potential to investigate two families of configurations of the magnetic field: one with, and one without, a field reversal, for a variety of nominal momenta and beta functions.

---

[†] p.drumm@rl.ac.uk, for the MICE collaboration.

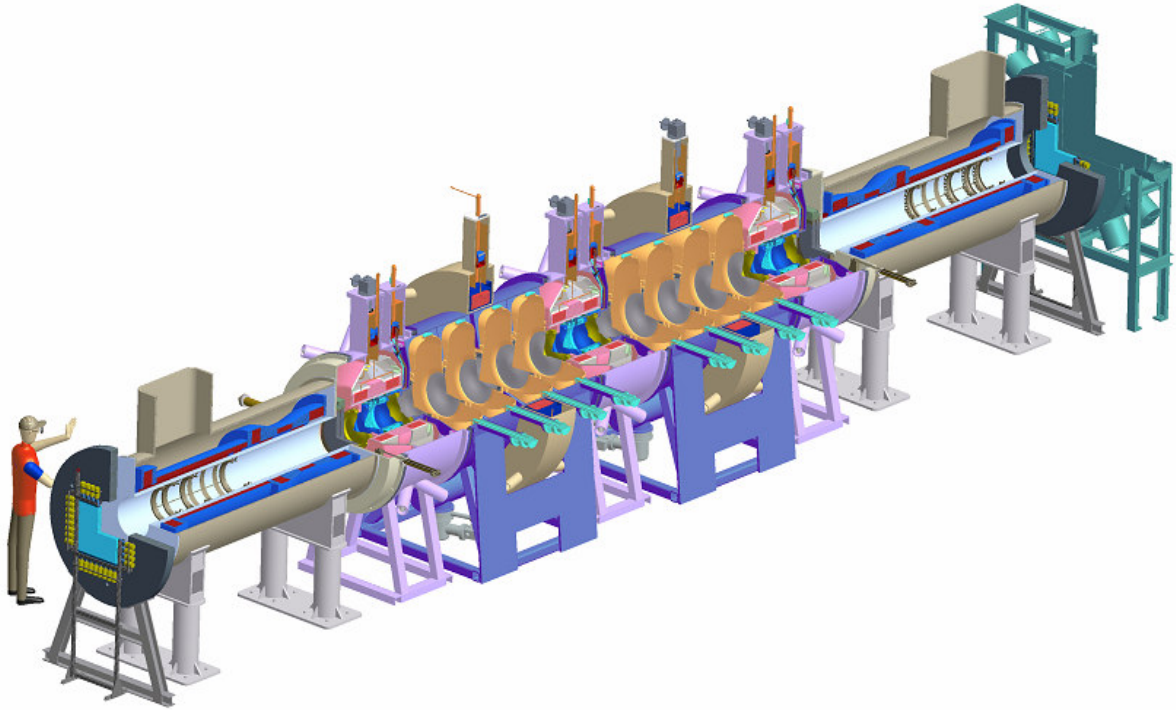

Figure 3. Schematic of the MICE cooling channel showing (from left to right) the TOF detector, spectrometer module, the absorber focus coil module and the cavity and coupling coil module. Additional TOF detectors, Cherenkov and Calorimeter detectors are shown at the far end. The channel has a solenoidal magnetic field along its length.

### *The Focus Coil and Absorber Module*

Hydrogen as a liquid in the absorber presents a considerable safety challenge. Significant effort has been expended to produce a robust and safe design. The system incorporates cryo-coolers and is designed to have a minimal heat leak [3]. A metal hydride is used to store the hydrogen when the absorber is emptied. An R&D program is being pursued to optimise the hydrogen [4] and absorber system designs [5].

### *RF Cavity and Coupling Coil Module*

The large cavities needed for MICE are being developed by the MUCOOL collaboration [6]. The influence of the combination of strong RF and static magnetic fields on the level of field emission is a serious challenge for a NF design [7]. For MICE, the requirements are still challenging but should be achievable.

RF power is provided by amplifiers from CERN and LBNL, with a baseline power level of >1MW to each of the eight cavities.

### *Putting it together*

The cooling channel modules are designed to sustain the magnetic forces arising from both operational and fault conditions. The internal forces of each magnet system are contained internally. External forces are taken through the body of the complete assembly, resulting in zero net external force on the support structure. Accurate alignment is not expected to be critical. Simulations with random 1 mm misalignments do not indicate any performance degradation. Nonetheless, best engineering practice is adopted throughout and alignment tolerances of 0.1 mm locally, and ~1 mm overall are expected.

## INSTRUMENTATION

Instrumentation for the MICE channel has received important consideration.

### *Particle Identification and time measurement*

The muon beam may contain residual pions which are transported through the large momentum acceptance of the beam line, and electrons from the in-flight decay of muons. A three-plane time-of-flight system provides the precise time information for emittance measurement and particle-identification. Additional particle identification is provided before and after the cooling channel by Cherenkov detectors and a final calorimeter.

### *Particle Tracking Detector*

A tracking detector has been constructed from planes of multiple layers of scintillating fibres. These planes, which sit in a uniform solenoidal field, provide position information at five points along the helical particle path that can then be reconstructed using pattern recognition techniques. The tracking detector provides a measure of the incoming and outgoing position, angles and momentum of each particle.

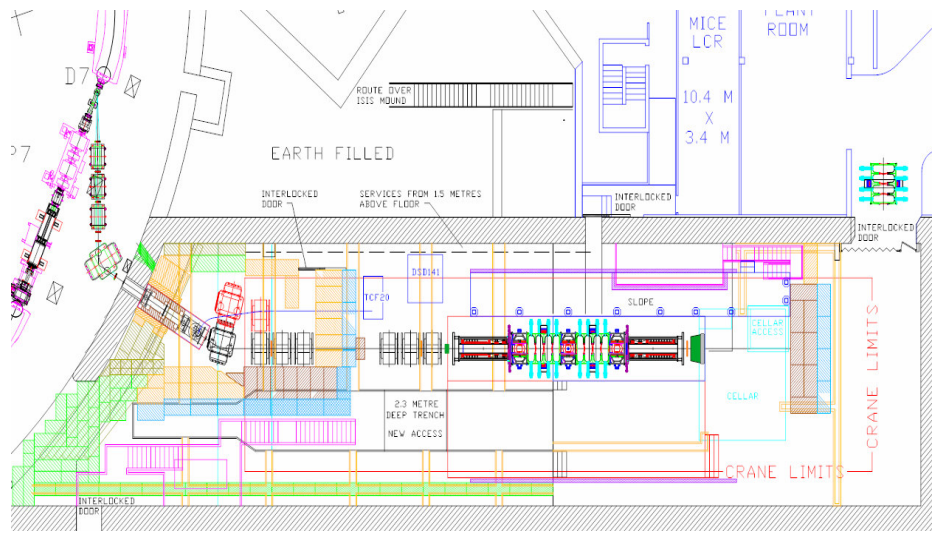

Figure 4. Schematic of the layout of MICE and the ISIS synchrotron.

*Estimated system performance*

Initial performance estimates have been made using a Monte Carlo model of the cooling channel, the detection systems and the beam. Although not complete, the performance of the channel in terms of the beam optics and of the detector performance requirements has been shown to meet the $10^{-3}$ precision specification.

*Environmental information*

MICE measurements include precisely determined magnetic fields and RF amplitudes and phases. The magnetic field of each magnet will be carefully mapped. Calibration and on-line monitoring will be done using 3D hall probes. The RF phase and amplitude will be measured, digitised and stored with the detector information.

## IMPLEMENTATION AT RAL

ISIS is a spallation neutron source powered by an 800 MeV, 300 µA proton synchrotron that cycles at 50 Hz. A target will be installed in the synchrotron to clip the halo of the beam and feed the MICE beam line.

*The beam line*

A fairly conventional pion decay channel will be implemented to provide the muon beam, making use of existing magnets at RAL and a superconducting decay solenoid from the PSI laboratory. Momentum selection prior to, and after, decay helps to make a fairly clean muon beam with a large emittance and momentum spread. The beam is transported to MICE through a set of quadrupoles that match the beam to the acceptance of the cooling channel. An adjustable emittance is achieved by focusing the beam onto a thick lead plate "diffuser" in the first spectrometer solenoid.

*Infrastructure*

Besides standard infrastructure (cooling water, compressed air, survey, etc.), of particular note are the hydrogen systems and the support structure; the latter is needed to carry the forces that may result under off-normal conditions.

Using cryo-coolers for all new magnets and for the liquid ($H_2$ or He) absorbers provides a considerable cost advantage compared with using a central cryogenic plant because of the large incremental cost of such plant. A small cryogenic system is still required for the existing decay solenoid.

*Timescale*

The beam and infrastructure for the first phase of MICE are expected to be complete by the end of spring 2007. Further details of MICE can be found on the MICE home page [8].

## REFERENCES

[1] Feasibility Study-II of a Muon-Based Neutrino Source, ed., S. Ozaki, R. Palmer, M. Zisman, and J. Gallardo, BNL-52623 (2001). http://www.cap.bnl.gov/mumu/studyii/final_draft/chapter-5/chapter-5.pdf
[2] C. Rogers et al., MPPE013, PAC2005, Knoxville, these proceedings.
[3] M. Green et al., TPP018, these proceedings.
[4] See T. Bradshaw. MICE Collaboration meeting, October 2004. http://hep04.phys.iit.edu/cooldemo/cm/cm10/26.html
[5] See S. Ishimoto, MUTAC review, LBNL, April 2005, http://www.cap.bnl.gov/mumu/conf/MUTAC-050425/MUTAC-agenda-2005.html.
[6] R. Rimmer et al., TPPT029, these proceedings.
[7] J. Norem et al., WPAT029, these proceedings.

[8] see http://www.mice.iit.edu/, for the mice proposal and the mice technical reference